\documentclass[12pt]{article}
\usepackage{graphicx}
\newcommand{\be}{\begin{equation}}
\newcommand{\ee}{\end{equation}}
\newcommand{\ba}{\begin{eqnarray}}
\newcommand{\ea}{\end{eqnarray}}

\renewcommand{\kappa}{{k}}

\catcode`\@=11

\@addtoreset{equation}{section}
\def\theequation{\thesection\arabic{equation}}

\def\@normalsize{\@setsize\normalsize{15pt}\xiipt\@xiipt
\abovedisplayskip 14pt plus3pt minus3pt%
\belowdisplayskip \abovedisplayskip
\abovedisplayshortskip  \z@ plus3pt%
\belowdisplayshortskip  7pt plus3.5pt minus0pt}
\def\small{\@setsize\small{13.6pt}\xipt\@xipt
\abovedisplayskip 13pt plus3pt minus3pt%
\belowdisplayskip \abovedisplayskip
\abovedisplayshortskip  \z@ plus3pt%
\belowdisplayshortskip  7pt plus3.5pt minus0pt
\def\@listi{\parsep 4.5pt plus 2pt minus 1pt
            \itemsep \parsep
            \topsep 9pt plus 3pt minus 3pt}}

\def\underline#1{\relax\ifmmode\@@underline#1\else
        $\@@underline{\hbox{#1}}$\relax\fi}
\@twosidetrue \relax

\catcode`@=12

\catcode`\@=11

\def\section{\@startsection{section}{1}{\z@}{3.5ex plus 1ex minus
   .2ex}{2.3ex plus .2ex}{\large\bf}}
\def\thesection{\arabic{section}.}

\def\ps@headings{\def\@oddfoot{}\def\@evenfoot{}
\def\@oddhead{\hbox{}\hfill
        \makebox[.5\textwidth]{\raggedright\ignorespaces --\thepage{}--
        \hfill }}
\def\@evenhead{\@oddhead}
\def\subsectionmark##1{\markboth{##1}{}}
}

\ps@headings

\catcode`\@=12

\relax

\def\figcap{\section*{Figure Captions\markboth
        {FIGURECAPTIONS}{FIGURECAPTIONS}}\list
        {Fig. \arabic{enumi}:\hfill}{\settowidth\labelwidth{Fig. 999:}
        \leftmargin\labelwidth
        \advance\leftmargin\labelsep\usecounter{enumi}}}
 \relax
\def\tablecap{\section*{Table Captions\markboth
        {TABLECAPTIONS}{TABLECAPTIONS}}\list
        {Table \arabic{enumi}:\hfill}{\settowidth\labelwidth{Table 999:}
        \leftmargin\labelwidth
        \advance\leftmargin\labelsep\usecounter{enumi}}}
 \relax
\def\reflist{\section*{References\markboth
        {REFLIST}{REFLIST}}\list
        {[\arabic{enumi}]\hfill}{\settowidth\labelwidth{[999]}
        \leftmargin\labelwidth
        \advance\leftmargin\labelsep\usecounter{enumi}}}
 \relax

\catcode`\@=11

\def\marginnote#1{}
\newcount\hour
\newcount\minute
\newtoks\amorpm
\hour=\time\divide\hour by60 \minute=\time{\multiply\hour by60
\global\advance\minute by- \hour}
\edef\standardtime{{\ifnum\hour<12 \global\amorpm={am}%
    \else\global\amorpm={pm}\advance\hour by-12 \fi
    \ifnum\hour=0 \hour=12 \fi
    \number\hour:\ifnum\minute<100\fi\number\minute\the\amorpm}}
\edef\militarytime{\number\hour:\ifnum\minute<100\fi\number\minute}
\def\draftlabel#1{{\@bsphack\if@filesw {\let\thepage\relax
  \xdef\@gtempa{\write\@auxout{\string
    \newlabel{#1}{{\@currentlabel}{\thepage}}}}}\@gtempa
    \if@nobreak \ifvmode\nobreak\fi\fi\fi\@esphack}
     \gdef\@eqnlabel{#1}}
\def\@eqnlabel{}
\def\@vacuum{}
\def\draftmarginnote#1{\marginpar{\raggedright\scriptsize\tt#1}}
\def\draft{\oddsidemargin -.5truein
        \def\@oddfoot{\sl preliminary draft \hfil
        \rm\thepage\hfil\sl\today\quad\militarytime}
        \let\@evenfoot\@oddfoot \overfullrule 3pt
        \let\label=\draftlabel
        \let\marginnote=\draftmarginnote

\def\@eqnnum{(\theequation)\rlap{\kern\marginparsep\tt\@eqnlabel}%
\global\let\@eqnlabel\@vacuum}  }
\def\preprint{\twocolumn\sloppy\flushbottom\parindent 1em
        \leftmargini 2em\leftmarginv .5em\leftmarginvi .5em
        \oddsidemargin -.5in    \evensidemargin -.5in
        \columnsep 15mm \footheight 0pt
        \textwidth 250mmin      \topmargin  -.4in
        \headheight 12pt \topskip .4in
        \textheight 175mm
        \footskip 0pt

\def\@oddhead{\thepage\hfil\addtocounter{page}{1}\thepage}
        \let\@evenhead\@oddhead \def\@oddfoot{} \def\@evenfoot{}
}
\def\titlepage{\@restonecolfalse\if@twocolumn\@restonecoltrue\onecolumn
     \else \newpage \fi \thispagestyle{empty}\c@page\z@
        \def\thefootnote{\fnsymbol{footnote}} }
\def\endtitlepage{\if@restonecol\twocolumn \else  \fi
        \def\thefootnote{\arabic{footnote}}
        \setcounter{footnote}{0}}  
\catcode`@=12 \relax

\def\ps@headings{\def\@oddfoot{}\def\@evenfoot{}
\def\@oddhead{\hbox{}\hfill
        \makebox[.5\textwidth]{\raggedright\ignorespaces --\thepage{}--
        \hfill }}
\def\@evenhead{\@oddhead}
\def\subsectionmark##1{\markboth{##1}{}}
}

\ps@headings

\relax

\begin{document}
\begin{titlepage}

\begin{centering}
\begin{flushright}
hep-th/0107124 \\ July 2001
\end{flushright}

\vspace{0.05in}

{\Large {\bf Cosmological Evolution in a Type-$0$ String Theory }}

\vspace{0.05in}

{\bf G.~A.~Diamandis } and {\bf B.~C.~Georgalas}
\\
{\it Physics Department, Nuclear and Particle Physics Section,
University of Athens,\\ Panepistimioupolis GR 157 71, Ilisia,
Athens, Greece.} \\

\vspace{0.05in}

{\bf N.~E.~Mavromatos } \\ {\it Department of Physics, Theoretical
Physics, King's College London,\\ Strand, London WC2R 2LS, United
Kingdom.} \\

\vspace{0.05in} {\bf E.~Papantonopoulos} and {\bf I.~Pappa} \\ {\it
Department of Physics, National Technical University of Athens,\\
Zografou Campus GR 157 80, Athens, Greece.} \\

\vspace{0.1in}
 {\bf Abstract}

\end{centering}

\vspace{0.05in}

{ We study the cosmological evolution of a type-$0$ string theory
by employing non-criticality, which may be induced by 
fluctuations of the D3 brane worlds.
We check the consistency of the approach 
to ${\cal O}(\alpha ')$ in the corresponding $\sigma$-model. 
The ten-dimensional theory is reduced to an effective
four-dimensional model, with only time dependent fields. We show that the
four-dimensional universe has an inflationary phase and graceful
exit from it, while the other extra dimensions are stabilized to a
constant value, with the fifth dimension much larger than the
others. We pay particular attention to demonstrating the r\^ole of
tachyonic matter in 
inducing these features. The Universe asymptotes, for large times, to a 
non-accelerating linearly-expanding Universe with a 
time-dependent dilaton
and a relaxing to zero vacuum energy a l\'a quintessence.}

\vspace{0.1in}
\begin{flushleft}
\end{flushleft}

\end{titlepage}

\newpage
\section{Introduction}

Interest in type-$0$ string theories has recently arisen due to
their connection to four-dimensional $SU(N)$ gauge theory via
appropriate dualities~\cite{type0}. The type-$0$ string model
contains electrically-charged three branes, $N$ of which may be
stuck on top of each other to construct the dual of a gauge
theory. An interesting feature of these non-supersymmetric
theories is the appearance of tachyonic matter excitations. 
These, 
however, do not lead to an instability, since they couple to the
appropriate five-form field strength that drives the tachyon to
a non-trivial vacuum expectation value capable of reversing the
original (negative) sign of the mass-squared.

Irrespective of the connection with dual gauge theories, the
presence of tachyonic matter in type-$0$ string theories has
important consequences in implying a cosmological evolution of the
brane of inflationary type~\cite{papa}. The r\^ole of tachyonic
matter in inducing non-trivial cosmological evolution in
two-dimensional string models has been studied
previously~\cite{diamandis}, with interesting conclusions on the
possibility of inflationary phase  as well as `graceful' exit from
it~\cite{grace}. It should be stressed, however, that the graceful
exit issue has been resolved  only upon invoking non-criticality
of the underlying string theory, which is arguably consistent with
the  non-equilibrium nature of the inflationary phase. An
important issue in such non-critical string analyses is the
identification of target time with the Liouville
mode~\cite{emn}, the consistency of which with the presence of
an inflationary phase and the eventual graceful exit is a highly
non-trivial issue.

Such a consistency is equivalent to demonstrating the 
existence 
of a solution of a system of equations
which represent a generalization of the conformal invariance
conditions in the case of a non-critical Liouville string, as a
result of the restoration of conformal invariance after Liouville
dressing~\cite{ddk}. An interesting physical issue is what causes
the non-criticality of the type-$0$ string. At present, the best
scenario we can offer are the quantum fluctuations of the three
brane solitonic structures \cite{kmw,szabo}. In General Relativity
there are no Rigid Objects, due to general covariance, and so
recoil fluctuations of the brane worlds are expected in general.
Moreover reconciliation with quantum mechanics requires the
incorporation of position and momentum uncertainties of the brane
worlds.

Under the identification of the Liouville mode with the target
time, the induced non-criticality leads to interesting predictions
on the possibility of an inflationary phase of the fluctuating
brane~\cite{adrian+mavro99}. However, the interesting issue is the
exit from such a phase. It appears that, apart from the graviton and
dilaton fields on the brane, the issue of graceful exit 
from the inflationary phase requires
the presence of additional matter excitations. It is the purpose
of this work to analyse an effective theory on a three-brane in
the context of a  non-critical type-$0$ string theory, with the
aim of demonstrating such a graceful exit from the inflationary phase. 
As
mentioned above, the non-criticality will be associated with
quantum fluctuations of the embedded three-branes.

We consider a non-critical ten-dimensional string theory of type-0,
and we
dimensionally reduce the ten-dimensional 
${\cal O}(\alpha ')$ string-effective action to
a four-dimensional space-time on a 3-brane world, assuming that all 
fields depend only upon time. We choose two different fields
to parametrize the internal space. The first field sets the scale
of the fifth dimension, while the other parametrize a flat five
dimensional space. In the same way, we dimensionally reduce the
modified (because of the off-criticality) $\beta$-functions of
the ten-dimensional theory, to the effective four-dimensional
$\beta$-functions, assuming a Robertson-Walker form of the
four-dimensional metric. Liouville dressing of the 
non-critical theory should restore conformal invariance
of the $\sigma$-model, which results in a set 
of generalized conformal invariance conditions. Upon the 
identification of the Liouville mode with the target time,
which eliminates scenaria with two target time-like coordinates, 
such conditions 
become equivalent to 
`equations of motion' for the appropriate fields.

These equations, 
supplemented by the Curci-Paffuti equation,
stemming from world-sheet renormalizability
of the $\sigma$-model,  
are studied in detail.  
At present, they appear to be 
too complicated to
be solved analytically. We thus follow a systematic method, 
that of
quasi-linear systems, to solve these equations numerically.

The main results of such an analysis may then be summarized 
as follows: 
our model clearly exhibits an
inflationary phase and, most importantly, a graceful exit from it. 
The supercriticality, as well as the tachyonic matter, 
play a crucial r\^ole to this effect. 
Moreover, 
there is stabilization of the internal space.
The
fifth dimension contracts during inflation and then freezes to
a constant value, while the other five internal dimensions also freeze
to a 
constant value, which 
in the solution we present in section four, 
is different from the first, 
indicating the possibility of obtaining different scales for the 
extra dimensions.

The structure of the article is as follows: in section two
 we give a brief review of the Liouville formalism under the
 identification of the Liouville mode with the target time.
 In section three we discuss the type-0 string theory and we
 derive the equations of motion of the effective non-critical
 four-dimensional theory. In section four, we describe the method for
 solving numerically the system of the differential equations
 describing the equations of motion and we support our findings by
 analytic arguments on the asymptotic behaviour of the solutions.
 We present the numerical results
and we discuss their physical significance. Finally, in section five, we
present our conclusions and outlook.
Some important formal properties of the dilaton field in the 
context of Liouville strings are discussed in an appendix. 

\section{Liouville String Formalism}

We commence our analysis with a brief review of the Liouville
dressing procedure for non-critical strings, with the Liouville
mode viewed as a local world-sheet renormalization group
scale~\cite{emn}. Consider a conformal $\sigma$-model, described
by an action $S^*$ on the world-sheet $\Sigma$, which is deformed
by (non conformal) deformations $\int_{\Sigma}g^iV_id^2\sigma$,
with $V_i$ appropriate vertex operators.
\be
S_g = S^* + \int_{\Sigma}g^iV_id^2\sigma
\label{sigma}
\ee

The non-conformal nature of the couplings $g^i$ implies that their
(flat)world sheet renormalization group $\beta$-functions,
$\beta^i$, are non-vanishing. The generic structure of such
$\beta$-functions, close to a fixed point, $\{g^i = 0\}$ reads:
\be
\beta^i = (h_i - 2)g^i + c^i_{jk}g^jg^k + o(g^3).
\label{fixed}
\ee

In the context of Liouville strings, world-sheet gravitational dressing is
required. The ``gravitationally''-dressed couplings, $\lambda^i(g,t)$,
which from our point of
view correspond to renormalized couplings in a curved space,
read to $O(g^2)$~\cite{ddk,liouville}:
\be
\lambda^i(g,t) = g^ie^{\alpha_it} + \frac{\pi}{Q \pm
2\alpha_i}c^i_{jk}g^j g^kte^{\alpha_it} + O(g^3), \qquad
Q^2 = \frac{1}{3}(c-c^*) \nonumber
\label{renorm}
\ee
where $t$ is the (zero mode) of the Liouville mode, 
$c^*$ is a critical (fixed point) value of the central charge,
about which we perturb the theory,  
$Q^2$ is the central
charge deficit, and $\alpha_i$ are the gravitational anomalous dimensions:
\be
\alpha_i(\alpha_i + Q) = h_i - 2 \qquad {\rm for} \qquad  c \ge c^*
\nonumber \label{anom} \ee Below we shall concentrate exclusively
to the supercritical string case, $Q^2 \ge 0$, which from the
point of view of identifying the Liouville mode with target time,
corresponds to a Minkowskian signature spacetime manifold.
The fixed point value $c^*$ 
is normalized such that for superstrings we are
dealing with here, one has 
$Q^2=\frac{1}{2}(d-9)$, in the special 
case of superstrings with non critical space time dimensionality ($d$).

Due to the renormalization (\ref{renorm}), the critical-string
conformal invariance conditions, amounting to the vanishing of
flat-space $\beta$-functions, are now substituted by:
\be
{\ddot \lambda}^i + Q{\dot \lambda}^i = -\beta^i(\lambda)  \qquad
{\rm for}~ c \ge c^*. \nonumber \label{neweq} \ee where the
notation $\beta^i(\lambda)$ denotes flat-world-sheet
$\beta$-functions but with the formal substitution $g^i
\rightarrow \lambda^i(g,t)$. Note the minus sign in front of the
flat world sheet $\beta$-functions $\beta^i$ in (\ref{neweq}),
which is characteristic of the supercriticality of the
string~\cite{ddk,liouville}. As we see later, the sign will be
crucial for the existence of acceptable inflationary solutions
demonstrating graceful exit from the exponential expansion phase.
Upon the identification of the Liouville mode with the target time
the dot denotes temporal derivative.

A highly non-trivial feature of the $\beta^i$ functions is the
fact that they are expressed as gradient flows in theory
space~\cite{zam,osborn}, i.e. there exists a `flow' function
${\cal F} [g]$ such that
\be\label{flow} \beta^i = {\cal
G}^{ij}\frac{\delta F[g]}{\delta g^j}
\ee
where ${\cal G}^{ij}$ is
the inverse of the Zamolodchikov metric in theory
space~\cite{zam}, which is given by appropriate two-point
correlation functions between vertex operators $V^i$. In the case
of stringy $\sigma$-models the flow function ${\cal F}$ may be
identified~\cite{osborn} with the running central charge deficit
$Q^2=c[g]-c^*$, where $c[g]$ is the central charge of the deformed
theory, and $c^*$ one of its critical values (conformal point)
about which the theory is perturbed by means of the operators
$V^i$.

It is then interesting to notice that, upon the identification of
the Liouville mode with the target time~\cite{emn}, the form of
equations (\ref{neweq}), upon the condition (\ref{flow}), is
reminiscent of inflaton equations in standard Cosmology, with the
r\^ole of the inflaton potential being played by the flow function
$Q^2[g]$ in theory space, and that of the Hubble parameter $H$ by
its square root $Q[g]$. There is an important difference, though,
as compared with the inflaton case, in the sense that here  all
the field modes seem to exhibit this behaviour.

An important comment we would like to make concerns the
possibility of deriving the set of equations (\ref{neweq}) from a
target space action. This issue has been discussed in the
affirmative in \cite{ emninfl}, where it was shown that the set of
equations (\ref{neweq}) satisfies the Helmholtz conditions for the
existence of an action in the `space of couplings' $\{ g^i \}$ of
the non-critical string. The property (\ref{flow}) is crucial to
this effect. Upon the identification of target time with the
Liouville mode~\cite{emn} this action becomes identical with the
target space action describing the off-shell dynamics of the
Liouville string. We should stress the fact that the action is off
shell, in the sense that the on-shell conditions correspond to the
vanishing of the $\beta$-functions $\beta^i$. In our case $\beta^i
\ne 0$, and the identification of the Liouville mode with the
target time implies that the space time graviton $\beta$-function
on the right hand side of (\ref{neweq}), as well as other
target-space tensorial structures, appearing inside the $\beta ^i$
functions for the various modes, contain temporal (Liouville)
components as well.

In this respect, our non-equilibrium Liouville string approach to
the temporal evolution should be contrasted with the
interpretation of a Liouville string as a critical equilibrium
string living in a space time with one extra  dimension. In that
case the corresponding $\beta$-functions of the Liouville-dressed
theory would satisfy classical equations of motion. As mentioned
above, in our approach the conditions describing  the restoration
of conformal invariance by means of Liouville dressing are {\it
not} to be interpreted as classical equations of motion of a
string living in a space-time with one extra target dimension.
Thus our analysis below should be distinguished from previous
analysis on Liouville cosmology~\cite{schmid}.

A generic feature of Liouville dynamics, is that in terms of the world-sheet
action, the normalization of the Liouville kinetic term can always be
arranged (by choosing appropriate counterterms) to correspond to a target
spacetime of Friedman-Robertson-Walker (FRW) type; i.e. the time-like metric
component (under the assumption that the Liouville mode is time) is:
\be
G_{00} = -1. \label{frwform} \ee We remind the reader that the
Minkowskian signature is due to supercriticality ($3Q^2 = c-c^* \ge
0$) assumption. This will be understood in what follows.

\section{Non-Criticality Induced in Type-$0$ String Theory}

We will start our discussion with the ten-dimensional type-$ 0$
string theory. The effective ten-dimensional target space action
of the type-$0$ String has the form, to ${\cal O}(\alpha ')$ in
the Regge slope $\alpha '$~\cite{type0}:
 \begin{eqnarray}\label{actiontype0} S&=&\int d^{10} x
\sqrt{-G}\Big{ [}e^{-2\Phi}\Big{(}R + 4(\partial_M \Phi)^2
-\frac{1}{4}(\partial_M T)^2 - \frac{1}{4}m^2T^2 \nonumber \\&~~ &
-\frac{1}{12}H_{MNP}^2\Big{)} -  \frac{1}{4}(1 + T +
\frac{T^2}{2})|{\cal F}_{MNP\Sigma T}|^2\Big{]} \end{eqnarray}
where capital Greek letters denote ten-dimensional indices, $\Phi$
is the dilaton, $H_{MNP}$ denotes the field strength of the
antisymmetric tensor field, which we shall ignore in the present
work, $T$ is the tachyon field of mass $m^2 <0$. In our analysis
we have ignored higher-order terms in the tachyon potential. The
quantity ${\cal F}_{MNP\Sigma T}$ denotes the appropriate
five-form of type-$0$ string theory, which couples to the tachyon
field in the Ramond-Ramond (RR) sector via the function $f(T)=1 +
T + \frac{1}{2}T^2$.

From (\ref{actiontype0}) one sees easily the important r\^ole of
the five-form ${\cal F}$ in stabilizing the ground state. Due to
its special coupling with the quadratic $T^2$ term in RR sector of
the theory, it yields an effective mass term for the tachyon which
is positive, despite the originally negative $m^2$
contribution~\cite{type0}.

To go off criticality we add the following term to the action
(\ref{actiontype0}) \ba \label{qterm}
 - \int d^{10} x
\sqrt{-G}e^{-2\Phi} Q^2 
\ea 
The sign of $Q^2$  is positive if one assumes
supercriticality of the string, which we shall do
here~\cite{aben,ddk,grace}. 
There may be various reasons for having such a term in the effective 
action. One concrete example of induced super-criticality 
in brane-world scenaria is the presence of impulse action (recoil)
on the $D3$-world brane as a result of either quantum fluctuations or 
scattering with other heavy defects or (a macroscopic number of) 
closed string states~\cite{kmw}.
As argued in those works the impulse deformations of the pertinent 
$\sigma$-model are relevant from a world-sheet renormalization-group 
view point, and as such require Liouville dressing~\cite{ddk}. 
The Liouville
mode may be identified with the target time in a way 
consistent with the restoration of conformal invariance. 
In general, there 
may be other microscopic reasons for the induction
of non-criticality in string theory, all related to some sort of 
non-equilibrium processes, which are expected to play a 
dominant r\^ole in early Universe Cosmologies.  
For the purposes of the present article, therefore,  
we shall 
not specify further the precise microscopic origin 
of such non-criticality, apart from associating it generically
with fluctuations of the brane worlds. In this sense 
we may   
assume $Q$ to be a rather `phenomenological' parameter, to be determined
by consistency with the conformal invariance conditions
of the Liouville-dressed~\cite{ddk} $\sigma$-model. 
For our purposes 
we shall restrict ourselves to ${\cal O}(\alpha')$ string 
effective actions.

In general, $Q^2$ depends on the $\sigma$-model
backgrounds fields, being the analogue of Zamolodchikov's
$C$-function~\cite{zam}. Assuming renormalizability, implies
that any explicit time-dependence of $Q$ will be absorbed
in renormalized couplings $g^I$:
\be\label{rgeq}
  \frac{d}{dt} Q= \frac{\partial }{\partial t} Q + \beta^I \partial_I Q=0
\ee However, the true set of non-marginal couplings in string
theory is actually infinite. Moreover, in our case, there are non
conformal contributions from world-sheet boundaries as well,
expressing, for instance, recoil of the (quantum) fluctuating D3
brane world.

Formally, such fluctuations may be approximated semi-classically
by impulse-type  
logarithmic conformal deformations at a $\sigma$-model level~\cite{kmw},
which have recently been shown to exist also in theories
with world-sheet fermions~\cite{szabo}, such as the type-$0$ string theories 
under consideration~\cite{type0}.
Such contributions can only be computed approximately
at present, for weakly coupled strings~\cite{kmw,szabo,kanti98}, 
and thus
are not taken into account explicitly in the effective
`phenomenological' approach adopted here, where only the bulk
(closed-string sector) contributions will be considered in a
detailed manner. We denote such contributions by an upper case
small Latin index $\{ g^i \}$, in contrast with the upper case
capital index $g^I$ which denotes the complete set of
non-conformal deformations. 

In view of (\ref{rgeq})
this implies that in our case we should consider $Q(t,g^i)$
as depending explicitly on time (renormalization-group scale $t$),
in addition to its implicit dependence through the renormalized
couplings $g^i$. This explicit time dependence simply expresses
the contributions to $Q$ from the rest of the non-conformal modes.
Its form, then, as a function of time $t$,  should be
determined by demanding consistency
with the generalized conformal-invariance
conditions (\ref{neweq}) obtained for the problem at hand from the
action (\ref{actiontype0}). This is the approach we shall assume from now on.

To this end, we need first to consider the dimensional reduction
of the ten-dimensional action to the four-dimensional space-time
on the brane. This procedure is achieved by assuming the following
ansatz for the ten-dimensional metric:
\begin{equation}
G_{MN}=\left(\begin{array}{ccc}g^{(4)}_{\mu\nu} \qquad 0 \qquad 0 \\
0 \qquad e^{2\sigma_1} \qquad 0 \\ 0 \qquad 0 \qquad
e^{2\sigma_2} I_{5\times 5} \end{array}\right)
\label{metriccomp}
\end{equation}
where lower-case Greek indices are four-dimensional space time
indices, and $I_{5\times 5}$ denotes the $5\times 5$ unit matrix.
We have chosen two different scales for internal space. The field
$\sigma_{1}$ sets the scale of the fifth dimension, while
$\sigma_{2}$ parametrize a flat five dimensional space. In the
context of homogeneous cosmological models, we are dealing with here, the
fields $g_{\mu\nu}^{(4)}$ are of
Robertson-Walker type, and $\sigma_{i},~i=1,2$ are assumed to
depend on the time $t$ only.

By varying the effective action with respect to the five-form one
obtains the corresponding $\beta$-function:
\be
{\tilde \beta}_5 = \nabla_M\left[(1 + T + \frac{T^2}{2}){\cal
F}^{MNP\Sigma T}\right] \label{fiveform} \ee It is important to
notice that in our approach we consider as a source of
non-criticality for the underlying string theory only quantum
fluctuations of the D3 brane resulting in the non-vanishing of the
renormalization group $\beta$-functions of the string multiplet,
i.e. graviton, dilaton and tachyon fields. The deformation
corresponding to the five-form~\cite{type0} is assumed conformal.
This is associated with that fact that this field is related to
the RR electric flux of the D-branes, and as such is not affected
by the (recoil) quantum fluctuations. Moreover, for our purposes
in this work, which is to demonstrate an inflationary phase and
graceful exit from it, the essential r\^ole is played by the
string multiplet fields, with the five-form flux field serving the
sole purpose of tachyon stabilization. This implies the vanishing
of the left-hand side of (\ref{fiveform}):
\be
0 = \nabla_M\left[(1 + T
+ \frac{T^2}{2}){\cal F}^{MNP\Sigma T}\right]
\label{fiveform2}
\ee
Moreover, for simplicity we shall assume a configuration for which
the five-form acquires non-vanishing values only along one extra
dimension:
\be
{\cal F}_{MNP\Sigma T}=\varepsilon _{0 \alpha \beta \gamma \delta
}~f_5(t) \label{reduction}
\end{equation}
where the lower-case Greek letters denote four-dimensional
indices.

The ten-dimensional Weyl-anomaly coefficients (generalized
$\beta$-functions~\cite{tseytlinshore}) for the $\sigma$-model
corresponding to propagation in a type-$0$ string background read,
to ${\cal O}(\alpha ')$: \ba {\tilde \beta}^G_{MN}&=&\{ R_{MN} + 2
\Phi _{;MN} - \frac{1}{4}T_{;M}T_{;N} \} + \nonumber \\ &~&
\frac{1}{2}e^{2\Phi}f(T)\left[G_{MN}({\cal F}_{K\Lambda P \Sigma
T})^2 - 10 {\cal F}_{MK \Lambda P \Sigma}{\cal F}_N^{K \Lambda P
\Sigma}\right]~, \nonumber \\ {\tilde \beta}^\Phi &=& \{ -R +
4(\partial_M \Phi)^2 - 4\nabla ^2 \Phi + \frac{1}{4}(\partial _M
T)^2 + \frac{1}{4}m^2T^2 + Q^2\}~, \nonumber \\ {\tilde \beta}^T
&=& \{ -\nabla ^2 T + 2 T_{;M}\Phi_{;}^M + m^2~T \} +
2e^{2\Phi}f'(T)\left({\cal F}_{MNP\Sigma T}\right)^2
\label{betafunctions} \ea

Due to the renormalizability of the (non-critical) $\sigma$-model,
there is an additional equation~\cite{curci} which should
supplement equation (\ref{neweq}), the Curci-Paffutti relation,
which relates the dilaton $\beta$-function, and hence the
effective running central charge of the theory, with the rest of
the $\beta$-functions:
\be
-\nabla _M {\tilde \beta}^\Phi + 2 G^{NP}e^{2\Phi} \nabla _N
(e^{-2\Phi} {\tilde \beta}^G_{MP}) + T_{;M} {\tilde \beta}^T = 0
\label{pafuti} \ee  Although this equation holds formally in the
flat world sheet case, however in our framework it should also
hold for the $\beta^i(\lambda)$ functions, i.e. the flat world
sheet $\beta$-functions upon the substitution of the
$\sigma$-model couplings with the Liouville dressed ones. It
provides a highly non-trivial constraint~\cite{grace}, which
should be respected by the process of identifying the Liouville
(world-sheet) renormalization scale with the target
time~\cite{emn}. As emphasized by Tseytlin~ \cite{tseytlinshore}
this equation expresses, under the off-shell equivalence of the
Weyl-anomaly coefficients ${\tilde \beta}^i$ with the field
variations obtained from the target space string effective action,
the invariance of the latter under general coordinate
transformations. As such, the extension of this consistency
condition to our Liouville string case, with the Liouville mode
playing the r\^ole of the target time, is evident. Notice that,
due to the criticality condition (\ref{fiveform2}), the five-form
$\beta$-function does not appear in (\ref{pafuti}).

An important comment is now in order concerning the 
r\^ole of the dilaton field in non-critical string theories. 
In general, it is not clear that the dilaton 
can be treated as one of the other couplings $g^i$ 
of the deformed $\sigma$-model. 
In the Appendix 
we specify the precise conditions 
under which this can indeed be made possible, thereby allowing
the application of (\ref{neweq}) to the dilaton coupling as well.

Upon considering the fields to be time dependent only, 
i.e. considering spherically-symmetric homogeneous backgrounds, 
restricting
ourselves to the compactification (\ref{metriccomp}), and assuming
a Robertson-Walker form of the four-dimensional metric, with scale
factor $a(t)$, the generalized conformal invariance conditions
(\ref{neweq}), using (\ref{betafunctions}),(\ref{pafuti}) imply
the following equations: \ba &~& -3\frac{\ddot a}{a} + {\ddot
\sigma}_1 + 5 {\ddot \sigma}_2 - 2 {\ddot \Phi} + {\dot
\sigma}_1^2  + 5 {\dot \sigma}_2^2 + \frac{1}{4}{\dot T}^2 +
e^{-2\sigma _1 + 2\Phi}f_5^2 f(T)=0~, \nonumber \\ &~&  {\ddot a}a
+ a{\dot a}\left(2Q + {\dot \sigma}_1 + 5{\dot \sigma}_2 - 2{\dot
\Phi}\right)+ e^{-2\sigma_1 + 2\Phi}f_5^2 f(T)a^2 =0~, \nonumber
\\ &~& 3{\ddot \sigma}_1 + 5{\dot \sigma}_1^2 + 3\frac{{\dot
a}}{a}{\dot \sigma}_1 + 2Q{\dot \sigma}_1 + 5{\dot \sigma}_1{\dot
\sigma}_2 -  2{\dot \sigma}_1{\dot \Phi}+ e^{-2\sigma_1 +
2\Phi}f_5^2~f(T)=0 ~, \nonumber \\ &~&  3{\ddot \sigma}_2 + 9{\dot
\sigma}_2^2 + 3\frac{{\dot a}}{a}{\dot \sigma}_2 + 2Q{\dot
\sigma}_2 + {\dot \sigma}_1{\dot \sigma}_2 -  2{\dot
\sigma}_2{\dot \Phi}- e^{-2\sigma_1 + 2\Phi}f_5^2~f(T)=0~,
\nonumber
\\ &~&  2{\ddot T} + 3\frac{{\dot a}}{a}{\dot T} + Q~{\dot T} +
{\dot \sigma}_1 {\dot T} + 5{\dot \sigma}_2 {\dot T} - 2{\dot
T}{\dot \Phi} + \nonumber \\ &~& m^2T - 4 e^{-2\sigma_1 +
2\Phi}f_5^2 f'(T)=0~, \nonumber \\ &~&  {\ddot \Phi} + Q{\dot \Phi
} + 6\frac{{\dot a}}{a} + 6\frac{{\dot a}^2}{a^2} + \nonumber
\\ &~&2\left[-{\ddot \sigma}_1 - \frac{3{\dot a}}{a}\sigma_1
-5{\ddot \sigma}_2 - 15\frac{{\dot a}}{a}{\dot \sigma}_2 - {\dot
\sigma}_1^2 - 15 {\dot \sigma}_2^2 - 5 {\dot \sigma}_1{\dot
\sigma}_2 - \right.\nonumber \\ &~&2\left.{\dot \Phi}^2 + 2{\ddot
\Phi} + 6\frac{{\dot a}}{a}{\dot \Phi} + 2{\dot \sigma}_1{\dot
\Phi} + 10{\dot \sigma}_2{\dot \Phi} \right] - \frac{1}{4}{\dot
T}^2 + \frac{1}{4}m^2T^2 + Q^2=0~,\nonumber \\ &~&
C_{5}=e^{-\sigma_1 + 5 \sigma_2}f(T)f_5~, \nonumber \\ &~&
{\Phi}^{(3)} + Q{\ddot \Phi} + {\dot Q}{\dot \Phi} + 12\frac{{\dot
a}}{a^3}\left(a{\ddot a} + {\dot a}^2 + Q~a{\dot a}\right) -{\dot
T}({\ddot T}+ Q{\dot T}) + \nonumber \\ &~& 4{\dot
\sigma}_1({\ddot \sigma}_1 + 2{\dot \sigma}_1^2 + Q{\dot
\sigma}_1) + 20{\dot \sigma}_2({\ddot \sigma}_2 + 2{\dot
\sigma}_2^2 + Q{\dot \sigma}_2)=0 \label{eqsmotion} \ea where
$f'(T)$ denotes functional differentiation with respect to the
field $T$, the overdot denotes time derivative, and $\Phi^{(3)}$ denotes
triple time derivative.

Note that the above equations have been derived from a
ten-dimensional action. An equivalent set of equations (in fact at
most linear combinations) come out from the corresponding
four-dimensional action after dimensional reduction. Of course
this reduction leads to the string frame.
 We may turn to the Einstein frame through the
 transformation
 \begin{equation}
 g_E=e^{-2\Phi+\sigma _1 + 5\sigma _2}g
 \end{equation}
 In this frame the line element is
 \begin{equation}
 ds_E^2 =  -e^{-2\Phi+\sigma _1 + 5\sigma _2}dt^2
 +a^2(t) e^{-2\Phi+\sigma _1 + 5\sigma _2}(dr^2 +r^{2}d\Omega^{2})
 \end{equation}
Therefore to discuss cosmological evolution we
 have to pass to the cosmological time defined by
\begin{equation}\label{l1}
  e^{-\Phi+\frac{ \sigma _1 + 5 \sigma _2}{2}}dt = dt_E
\end{equation}\
Then the line element becomes
\begin{equation}\label{l2}
  ds_E^2 = -dt_E^2+a'^2(t_E)(dr^2+r^{2}d\Omega^{2})
\end{equation}
  with $a'(t_E)=
  e^{-\Phi+\frac{\sigma _1 + 5\sigma _2}{2}}a(t(t_E))$.

We should also remark that we have adopted the usual Kaluza-Klein
reduction without considering the extra dimensions compactified.
Nevertheless the equations we are interested in and the results we
will discuss in the following section are not modified if we had
considered a compact six-dimensional space instead. In that case
$e^{2\sigma_1}$ and $e^{2\sigma_2}$ should correspond to radii of
the compact space. Note also that in the string frame we have the
exponential $e^{-2\Phi+\sigma_1+5\sigma_2}$ instead of
$e^{-2\Phi}$, since we allow time dependence of the volume of the
compact space.

\section{Solution of the System of Differential Equations}

The problem of initial conditions of the Universe is 
a very complicated problem in Cosmology, and the present 
four-dimensional model is no exception. 
To bypass this problem
we adopt here a different strategy. 
We demand that the solution approach asymptotically
the Minkowski space time (in the sigma-model framework),
with a linear dilaton.
This space time is well-studied from the point of view 
of non-critical string theory~\cite{aben}, and provides
an example of a linearly expanding, non accelerating 
Universe in Einstein frame~\footnote{Concerning 
the model of \cite{aben} there has been some discussion
in the literature~\cite{sanchez} as to whether
this type of universe is really (physically) expanding or is 
actually static,
in case one accepts the point of 
view that the length measurements are made by string rods.
In our context there is no such ambiguity, because, as we shall
see below, this type of Universe is the final stage of 
the Liouville-time evolution, and for us the Einstein frame,
in which one recovers the linearly expanding Universe of \cite{aben}, 
is the physical frame.}. 
From the point of view of a local field theory, 
the fact that the acceleration of the type-$0$-string Universe 
is not eternal
is essential 
for a consistent quantization, and definition of
asymptotic states, and hence a scattering matrix~\cite{eternal,
emnsmatrix,Witten}.

We therefore assume that the universe of a type-$0$ non-critical
string theory is described by
a configuration,
which, in the 
$\sigma$-model frame, 
asymptotically (i.e. for large times) 
approaches a Minkowski spacetime with linear
dilaton and constant values of $\sigma_{1}$ and $\sigma_{2}$, 
and examine whether such an assumption
leads to a
consistent (and continuous) solution 
of the system of equations (\ref{eqsmotion})
for all times. 
For this purpose 
we separate
the fields in their asymptotic values plus fields which tend
asymptotically to zero. 
Substituting these fields back to the
equations we let the system evolve in time backwards.

We now describe the various steps of this procedure in detail. 
The 
separation of the various fields in their
asymptotic values plus fields which tend asymptotically to zero,
is performed as follows:  
\begin{eqnarray}\label{initval}
\Phi(t) &\equiv& f_{0}+f_{1}t+h(t)  \nonumber \\ a(t)& \equiv &
a_{0}e^{b(t)} \nonumber \\ \sigma_{1}(t) & \equiv &
s_{01}+s_{1}(t) \nonumber \\ \sigma_{2}(t) &\equiv &
s_{02}+s_{2}(t) \nonumber \\ Q(t) &\equiv & q_{0} +q_{1}(t)
\nonumber \\ T(t)&\equiv& c_{0}+{\cal T}_{0}(t) \nonumber \\ F_{5}(t)
&\equiv & e^{\Phi }f_{5}(t)
\end{eqnarray}
where the constants $q_{0}$ and $f_{1}$ are related through the
relation $q_{0}=-\frac{f_{1}}{2}\Big{\{} 1+\sqrt{17} \big{\}}$.
This relation results from the requirement that the dilaton
equation is satisfied. The fields $\Big{\{} h(t), b(t), s_{1}(t),
s_{2}(t), q_{1}(t), {\cal T}_{0}(t),f_{5}(t) \Big{\}}$ vanish
asymptotically. We want gravity to be weak asymptotically so we
choose $f_{1}$ to be negative. Due to the bare tachyonic
mass, $m^0<0$, $c_0$ must be chosen to vanish for the tachyon
equation to be satisfied asymptotically.
Note that the above asymptotic
conditions in the Einstein frame cosmology correspond to a linear
expanding universe \cite{aben}. With this choice the system of the
equations (\ref{eqsmotion}) become quasi-linear and, 
in order to reduce the
order of the system, we define
\begin{eqnarray}\label{dirinitval}
h(t)&\equiv&h_{0}(t) \nonumber \\ \dot{h}(t) &\equiv& h_{1}(t)
\nonumber \\ b(t)&\equiv&b_{0}(t)\nonumber \\ \dot{b}(t)& \equiv &
b_{1}(t) \nonumber
\\ s_{1}(t)&\equiv&s_{10}(t) \nonumber \\
\dot{s_{1}}(t) & \equiv & s_{11}(t) \nonumber \\
s_{2}(t)&\equiv&s_{20}(t) \nonumber \\  \dot{s_{2}}(t) &\equiv &
s_{21}(t) \nonumber \\ {\cal T}(t) &\equiv & {\cal T}_{0}(t) \nonumber \\
\dot{{\cal T}}(t)&\equiv& {\cal T}_{1}(t)
\end{eqnarray}
So the system of equations (\ref{eqsmotion}) take the form
 \begin{equation}\label{quasilinear}
 \dot{\vec{x}} =  {\bf A} \vec{x} + {\bf \vec{F}}(\vec{x})
 \end{equation}
where $\vec{x} = \Big{\{} h_{0}(t), h_{1}(t), q_{1}(t), b_{0}(t),
  b_{1}(t), s_{10}(t), s_{11}(t), s_{20}(t), s_{21}(t), {\cal T}_{0}(t), 
{\cal T}_{1}(t),
f_{5}(t) \Big{\}} ^{\bot}$, with $\bot $ denoting transpose, 
${\bf A}$ is the constant
matrix determining the linear part of the system and ${\bf
 \vec{F}}(\vec{x})$ gives the nonlinear terms.
 \newline
 The differential system (\ref{quasilinear}) can be written
 formally as an integral system
  \begin{equation}\label{itergalsol}
  \vec{x}(t) =  \vec{x}(t_{0}) + {\bf Y}(t) \int_{t_0}^{t}ds 
 {\bf Y}^{-1}(s) \vec{{\bf F}} \left[ \vec{x}(s) \right],
  \end{equation}
   where the matrix ${\bf Y}$ is a fundamental solution of the linear system
   satisfying the equation
 \begin{equation}
 \dot{{\bf Y}} =  {\bf AY} \nonumber
 \end{equation}
 The full solution of this integral system can be given in an iterative form:
 \begin{equation}\label{itersol}
  \vec{x}_{(n+1)}(t) =  \vec{x}_{(n)}(t) + {\bf Y}(t) \int_{t_0}^{t}ds
 {\bf Y}^{-1}(s) \vec{{\bf F}} \left[ \vec{x}_{(n)}(s) \right]
  \end{equation}

 The starting point of the iteration procedure is the solution of
 the linear system with the correct asymptotic behaviour.
If we insert the fields (\ref{initval}) and (\ref{dirinitval})
into the system of equations (\ref{eqsmotion}) and keep only the
linear part we get the following general solution of the system
\begin{eqnarray}
h(t) &=&  Ch_0 + Ch_{1}t + \frac{3}{4}
\frac{(Cs_{11}+5Cs_{21})}{(f_1 - q_0)}
e^{\frac{2}{3}(f_{1}-q_{0})t} \nonumber
\\  &-& \frac{3}{4}
\frac{Cb_{1}}{(f_1 - q_0)}e^{2 (f_{1}-q_{0})t} +
 \frac{C_{5}Cf_{5}(6f_{1}^2+9f_{1}q_{0}+q_{0}^2)e^{f_{0}-s_{01}-5s_{02}+2f_{1}t}}
 {8f_{1}q_{0}(2f_{1}+q_{0})}\nonumber
 \\
 q_{1}(t)
&=&
Cq_{1}+\frac{3Cb_{1}(2f_{1}-q_{0})}{2f_{1}}e^{2(f_{1}-q_{0})t}\nonumber
\\ &-&\frac{(Cs_{11}+5Cs_{21})(2f_{1}+q_{0})}{6f_{1}}e^{\frac{2}{3}(f_{1}-q_{0})t}
\nonumber \\  &-&\frac{C_{5}Cf_{5}(6f_{1}^2+9f_{1}q_{0}+q_{0}^2)}
 {8f_{1}^2q_{0}}e^{f_{0}-s_{01}-5s_{02}+2f_{1}t}
 \nonumber \\
  b(t) &=& Cb_{0} + \frac{Cb_{1}}{2(f_{1}-q_{0})}e^{2(f_{1}-q_{0})t}
 -\frac{C_{5}Cf_{5}}
 {8f_{1}q_{0}}e^{f_{0}-s_{01}-5s_{02}+2f_{1}t}  \nonumber \\
s_{1}(t) &=&  Cs_{10} +
\frac{3Cs_{11}}{2(f_{1}-q_{0})}e^{\frac{2}{3}(f_{1}-q_{0})t} -
\frac{C_{5}Cf_{5}}
 {8f_{1}(2f_{1}+q_{0})}e^{f_{0}-s_{01}-5s_{02}+2f_{1}t}
 \nonumber
\\
 s_{2}(t) &=&  C_{20} +
\frac{3Cs_{21}}{2(f_{1}-q_{0})}e^{\frac{2}{3}(f_{1}-q_{0})t} +
\frac{C_{5}Cf_{5}}
 {8f_{1}(2f_{1}+q_{0})}e^{f_{0}-s_{01}-5s_{02}+2f_{1}t}\nonumber \\
F_{5}(t)&=& Cf_{5}e^{f_{1}t}\nonumber \\ {\cal T}_{0}(t) &=& C_{1}e^{At}
+C_{2}e^{Bt}+ \frac{C_{5}Cf_{5}}{8f_{1}^2+2m^2+4f_{1}q_{0}}
e^{f_{0}-s_{01}-5s_{02}+2f_{1}t}
\end{eqnarray}
where
\begin{eqnarray}
A&\equiv&
\frac{1}{4}\Big{[}(2f_{1}-q_{0})\sqrt{(2f_{1}-q_{0})^{2}-8m^{2}}\Big{]}<0
\nonumber \\ B&\equiv&
\frac{1}{4}\Big{[}(2f_{1}+q_{0})\sqrt{(2f_{1}-q_{0})^{2}-8m^{2}}\Big{]}>0,
\end{eqnarray}
and $\Big{\{}Ch_{0}, Ch_{1}, Cq_{1}, Cb_{0}, Cb_{1}, Cs_{10},
Cs_{11}, Cs_{20}, Cs_{22}, C_{1}, C_{2}, Cf_{5}\Big{\}}$ are
constants of integration. To have the right asymptotic behaviour
we choose $\Big{\{}Ch_{0}, Ch_{1}, Cb_{0}, Cq_{1},Cs_{10},
Cs_{20}\Big{\}}$ to be zero. The constant $C_{2}$ corresponds to a
positive eigenvalue of the matrix ${\bf A}$, so we set it also
zero.
 Since the constant matrix ${\bf A}$ has negative eigenvalues,
  well known theorems from qualitative theory of quasi-linear
  systems, guarantee the asymptotic stability of the iterative solution
  \cite{book}

 Now  at the first step in the iteration
  the solution is given by:
 \be
 \vec{x}(t) = \vec{x}_{(0)}(t)  + {\bf Y}(t) \int_{t_0}^{t}ds 
 {\bf Y}^{-1}(s) \vec{{\bf F}} \left[{\bf Y}(s) \vec{C} \right],
 \label{firstiter}
 \ee
 where $\vec{x}_{(0)}(t)$ is given by,
 \begin{equation}
 \vec{x}_{(0)}(t)= {\bf Y}(t) \vec{C}
 \end{equation}
 and $\vec{C}$ is a constant vector $\vec{C}=\Big{\{}Ch_{0},
 Ch_{1}, Cq_{1}, Cb_{0}, Cb_{1}, Cs_{10},
Cs_{11}, Cs_{20}, Cs_{22}, C_{1}, C_{2}, Cf_{5}\Big{\}}$  with
  $Ch_{0}=Ch_{1}=Cq_{1}=Cb_{0}=Cs_{10}=Cs_{20}=C_{2}=0$
  for reasons we explained above.

The first step of the iteration can be integrated analytically and
has the right asymptotic behaviour as expected. Nevertheless due
to the complexity of the expressions and the fact that they do
not give us the full solution of the problem, we proceed as
follows. We take from this step the value of the fields at late
times and we use them as initial conditions for the numerical
solution of the system.

\begin{figure}[h]
\centering
\includegraphics[scale=0.9]{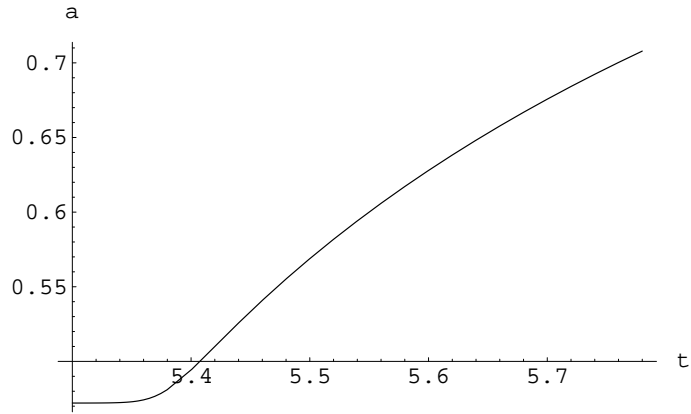}
\caption {The evolution of the scale factor in time.}
\end{figure}

We can now discuss our results. As we can see from Fig.1-Fig.3,
the cosmological evolution of our universe passes through the
following phases. At very early times, the universe starts from
the initial singularity. Then enters a phase where the physical
dimensions are formed. At this stage $\sigma_{1}$, $\sigma_{2}$
and $a$ are comparable in magnitude. The phase of inflation
follows, during which the scale factor grows exponentially (for a
short time though), while the internal space contracts with very
negative values of $\ddot{\sigma_{1}}$ and $\ddot{\sigma_{2}}$.
Finally, the universe enters a phase, where  
its expansion slows down
until it reaches the asymptotically flat space
(in the $\sigma$-model frame). The internal space
continues to contract but with very slow rate (with
$\ddot{\sigma_{1}}$ and $\ddot{\sigma_{2}}$ positive) until it
reaches a constant value. As we can see from Fig.2 and Fig.3,
$\sigma_{1}$ and $\sigma_{2}$ scale differently. The field
$\sigma_{2}$ very soon freezes to a value much lower than the
value of $\sigma_{1}$, indicating the the fifth dimension can be
much larger than the other five dimensions.

\begin{figure}[ht5]
\centering
\includegraphics[scale=0.9]{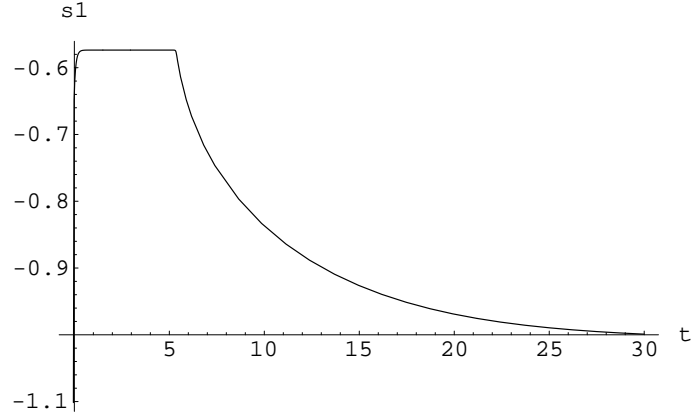}
\caption {The behaviour of the field representing the evolution of
the fifth dimension.}
\end{figure}

\begin{figure}[ht4]
\centering
\includegraphics[scale=0.9]{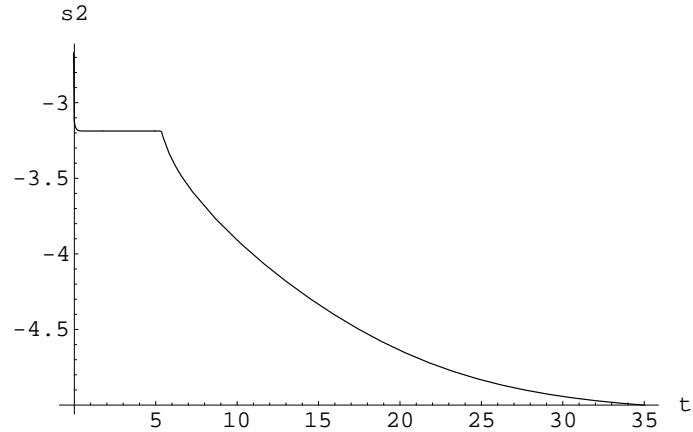}
\caption {The behaviour of the field $\sigma_2$ expressing the
evolution of the five internal dimensions. The difference of the
scales in the decomposition of the extra dimensions is clearly
seen.}
\end{figure}

For the dilaton field (Fig.4), we observe that at the singularity
the dilaton field is infinite, indicating that the gravity is very
strong. Then at the second phase of evolution, the strength of
gravity is weakened because the dilaton field drops linearly. 
This weakening of the gravitational 
interactions continues 
during inflation, and finally at
the exit, the dilaton field continues to drop linearly. The
Tachyon field (Fig.5) 
falls continuously during the evolution 
until it becomes zero. The RR-field (Fig. 6), 
on the other hand, grows 
from zero value at
the initial singularity until it reaches a constant value at the
exit of inflation.
Moreover, the behaviour of the central-charge deficit 
$Q^2$ is depicted in figure 7. 
The latter is initially decreasing, until it becomes zero, 
and then oscillates before reaching an asymptotic constant positive value 
$Q^2  \rightarrow {\rm const} \ne 0$, as $t \rightarrow \infty$.
A similar behaviour is observed in the two dimensional case~\cite{grace}.
Notice that the oscillations in the central charge deficit
are compatible with the fact that the Liouville mode is not a unitary 
$\sigma$-model field, as a result of its negative signature.
Hence, Zamolodchikov's $c$-theorem does not strictly apply,
although overall there is a reduction in the central charge,
since the theory starts with a very large value of $Q^2$ 
(formally infinite near the initial 
singularity), which  
eventually
asymptotes to a finite positive value.

\begin{figure}[ht3]
\centering
\includegraphics[scale=0.9]{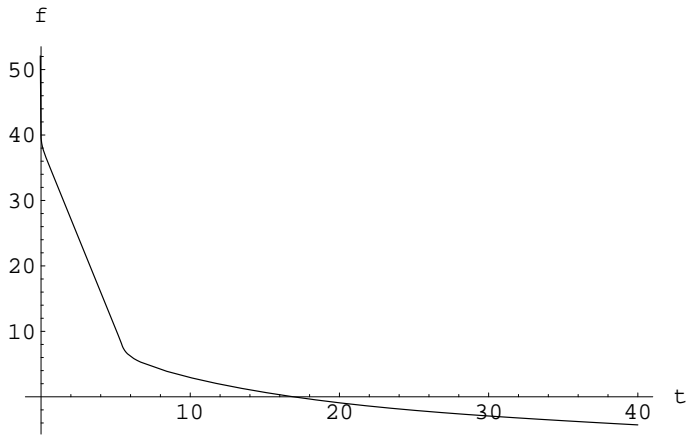}
\caption {The behaviour of the dilaton field.}
\end{figure}

\begin{figure}[ht2]
\centering
\includegraphics[scale=0.9]{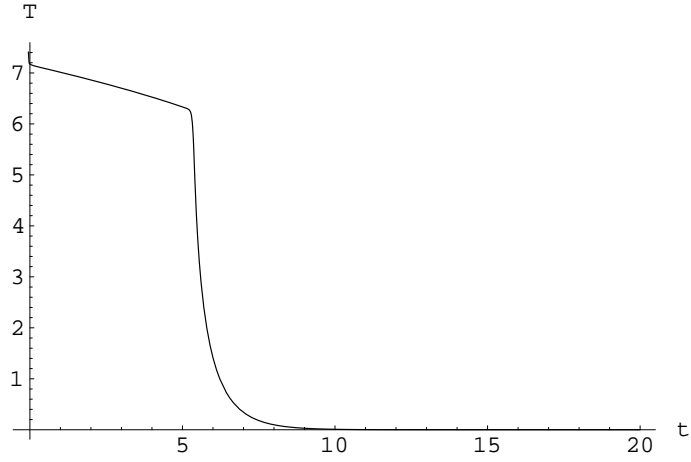}
\caption {The behaviour of the tachyon field.}
\end{figure}

\begin{figure}[ht1]
\centering
\includegraphics[scale=0.9]{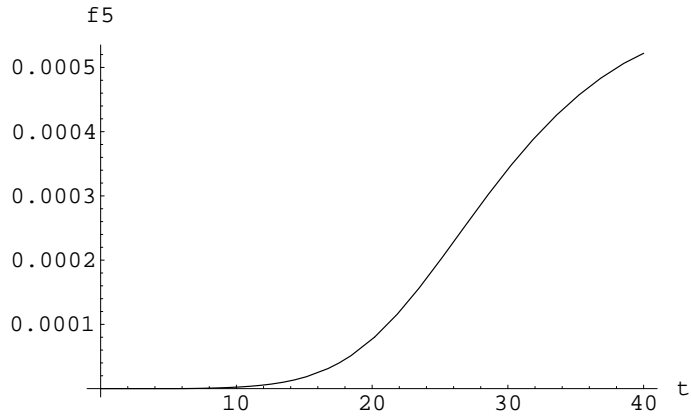}
\caption {The evolution of the field $f_5$ denoting the behaviour
of the R-R flux.}
\end{figure}

\begin{figure}[ht4]
\centering
\includegraphics[scale=0.9]{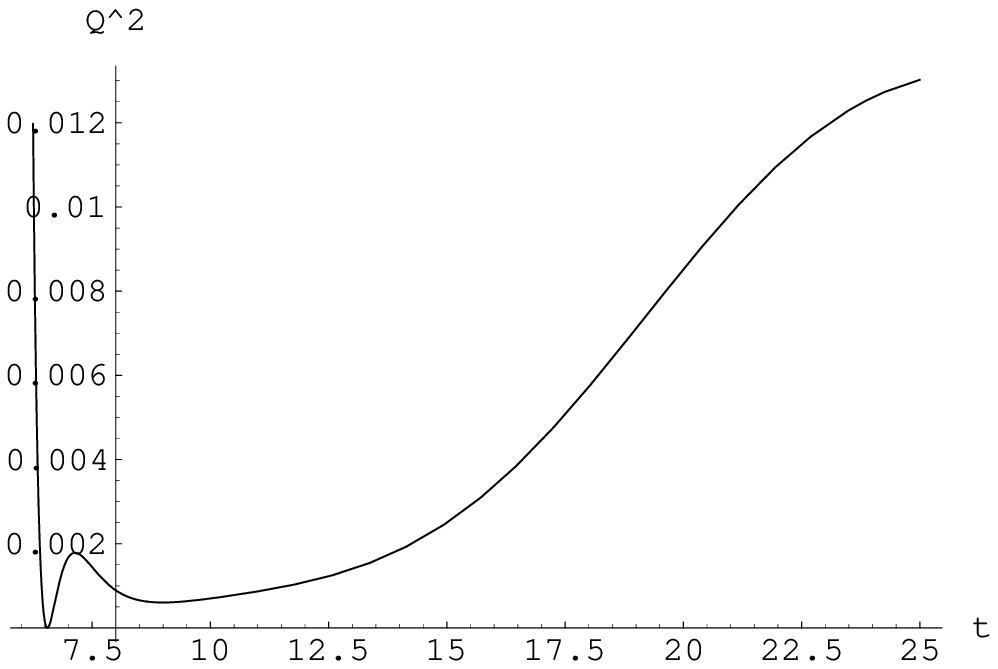}
\caption {The behaviour of the central charge deficit $Q^2$.
This quantity, 
after some oscillatory behaviour near its zero value,
asymptotes for large times to a constant value 
(not depicted
clearly in the figure  
due to numerical limitations).}
\end{figure}

We can analytically support the numerical solution we found
looking more carefully at the equations satisfied by the fields.
Assuming that asymptotically the fields $\sigma_{1}$ and
$\sigma_{2}$ are constant,  
using the second and seventh equation from
the system of equations (\ref{eqsmotion}), the field $b$ takes the
form of a Riccati equation, which is the usual case in
cosmological models based on type-0 string theory \cite{papa},
\begin{equation}\label{riccati}
\ddot{b}+\dot{b}\Big{(}2Q-2\dot{\Phi}+\dot{\sigma_{1}}
+5\dot{\sigma_{2}} \Big{)}+\dot{b}^{2}=-\frac{C_{5}}{2}
e^{-\sigma_{1}+5\sigma_{2}+\Phi}F_{5}
\end{equation}
For times nearly after the initial singularity and during
inflation, equation (\ref{riccati}) is approximated by
\begin{equation}\label{inflric}
\ddot{b}+2\dot{b}(Q(t)-\dot{\Phi})+\dot{b}^{2}=0
\end{equation}
This equation has the solution
\begin{equation}
b=c_1+t+ln(1-e^{-c_2t})
\end{equation}
where $c_1,c_2$ are constants. This solution shows that the scale factor
has an inflationary phase and for later times the logarithmic term
takes over and slows down the expansion of the universe.

 For large times when the internal dimensions have been stabilized
 in constant values, the
equation (\ref{riccati}) becomes
\begin{equation}\label{rrt}
\ddot{b}+2\dot{b}(q_{0}-f_{1})+\dot{b}^{2}=ce^{2f_{1}t}
\end{equation}
The exact solution of this equation respecting the required
asymptotic behaviour is
\begin{equation}\label{bequ}
b(t)=ln \left[ _0F_1 \left(
\frac{q_0}{f_1},\frac{c}{4f_1^2}e^{2f_1 t} \right) + c_3
{e^{2(f_1-q_0)t}} {_0F_1} \left(2-\frac{q_0}{f_1}, \frac{c}{4f_1^2}
e^{2f_1 t} \right) \right]
\end{equation}
where $c_{3}$ is a constant, and $_0F_1$ is the Confluent
Hypergeometric Function. If we expand this solution in series, we
get the asymptotic behaviour of $b(t)$ which characterizes the
inflationary exit period.
The quantity $Q(t)-\dot{\Phi}$ is
crucial for the existence of an inflationary phase 
and then exit from it. We note that in two
target dimensions, it is the same quantity that controls the exit from
inflation \cite{grace}.

We now make some brief remarks on the physical significance 
of the asymptotic behaviour of our solution
(for large times $t \rightarrow \infty$). 
A detailed discussion will appear
in a forthcoming publication~\cite{toappear}. 
By inspecting figure 1, or analytically from (\ref{bequ}), 
we observe that   
the acceleration of the universe
vanishes asymptotically in time ($t \rightarrow \infty$)
and one recovers a non-accelerating 
Minkowski universe with a linear dilaton 
in the $\sigma$-model (string) frame~\cite{aben}:  
${\ddot a}{a} = {\ddot b} + ({\dot b})^2 \rightarrow 0$, for $t \to \infty 
$, with $a(t) \to a_o$.

Such configurations imply the absence of a cosmological 
horizons, 
given that 
the relevant quantity diverges,  $\delta \propto \int _{t_0}^\infty dt/a(t) \rightarrow 
\infty$~\footnote{Notice that a divergent result for $\delta $ 
is, of course, also obtained in the 
Einstein frame (\ref{l1}), (\ref{l2}), when we express 
the integrand in terms 
of the Robertson-Walker time $t_E$. In that case
the scale factor is expanding  {\it linearly}
$a'(t_E) \propto t_E$, while the dilaton varies 
logarithmically with the time $t_E$.}.   
This is a welcome fact from the point of view 
of low-energy field theory, in the sense that 
in such not-eternally accelerating Universes 
one is able to define 
asymptotic states and thus a proper $S$ matrix~\cite{eternal}. 
Out of interest we notice that asymptotic situations of this kind 
appear to be generic in cosmologies
with fluctuating D-brane defects~\cite{kmw}.

Another issue that arises in this scenario is the asymptotic value of 
the vacuum energy. In the Einstein frame (\ref{l1}), 
the latter is given 
by 
\be
\Lambda = e^{2\Phi -\sigma_1 -5\sigma_2}Q^2 \rightarrow 
e^{-|const|\,t -\sigma_1- 5\sigma_2}Q^2 \rightarrow 0, \qquad t \to \infty   
\label{cosmol} \ee
where we took into account 
that 
asymptotically in time the fields $\sigma_{1},\sigma_2$ 
become constants, 
the dilaton configuration is linear and negative (see figure 4),
while the central-charge deficit $Q^2 \rightarrow {\rm const} > 0$.
In our solution, therefore, we observe that the 
four-dimensional vacuum energy is positive (de Sitter type) 
and tends to zero asymptotically. When we express the result (\ref{cosmol}) in 
terms of the Robertson-Walker time $t_E$ (\ref{l2}) 
we observe that the four-dimensional vacuum energy
relaxes to zero as: 
\be
\Lambda =({\rm const})\frac{1}{t_E^2}
\ee
which is compatible with recent experimental evidence, and 
is obtained in many quintessence models. 
In our approach the quintessence field is, in a sense, the 
dilaton~\cite{emninfl,emnsmatrix}.  
We shall present a more detailed account of these features in 
a forthcoming publication~\cite{toappear}.

We next remark that in the context of a critical string theory, 
with $Q=0$, the dilaton equation in (\ref{eqsmotion}) implies that 
the dilaton field $\Phi$ becomes asymptotically constant and, hence,
equation (\ref{rrt}) becomes
\begin{equation}
\ddot{b}+\dot{b} ^{2}=c
\end{equation}
This has as solution
\begin{equation}
b(t)=c_{1}+ln \Big{[}cosh \Big{\{}\sqrt{c}t-\sqrt{c}c_{2}\Big{\}}
\Big{]}
\end{equation}
where $c_{1}$ and $c_{2}$ are constants. Then the scale factor has
an exponential growth but it will expand for ever.
Such a case is inconsistent from the point of view of defining 
asymptotic states and an $S$-matrix. Because the case $Q=0$,
is supposed to correspond to critical string theory, 
which must be a theory of an $S$-matrix, we may conclude that 
such a situation will not arise as a consistent string theory solution.

All the analysis we have obtained so far pertains to the string 
($\sigma$-model) frame.
As already mentioned, one   
gets similar results in the Einstein frame,  using
(\ref{l1}) and (\ref{l2}).

Another consistency check of the approach is 
provided by the F-field of the RR-flux.
Because of equation (\ref{fiveform2}), the field $F_{5}(t)$ has
the exact solution
\begin{equation}
F_{5}(t)=e^{\phi}f_{5}=\frac{C_{5}e^{f_{0}+s_{01}-5s_{02}}}
{\Big{(}1+g(t)+\frac{g(t)^{2}}{2}\Big{)}} e^{f_{1}t+h(t)}
\end{equation}
Stability of the internal space and consistency of our scheme
requires the constant $f_{0}$ to have a large negative value. This
means that gravity is asymptotically very weak, as expected.

\section{Discussion}

In the present article 
we have presented a cosmological model based on a type-0 string
theory. The type-0 string theory is rich in its content. 
In addition to 
the graviton and dilaton fields, it includes also a tachyon
field which couples to an RR five-form field. To avoid tachyon
instabilities the tachyon field has to take on values at the minimum
of its potential. Considering the ten-dimensional action of the
type-0 field theory one can get at the conformal point the
$\beta$-functions of the theory, and from them,  
by reduction to four-dimensions,
and assuming that all fields are time dependent, one 
obtains an
effective four-dimensional theory. 
It is found that this theory, in the context of a
Robertson-Walker background, has an inflationary phase but 
there is no smooth 
exit from this phase.

We believe that the central issue of inflationary cosmology is not
how one can get an inflationary phase, but how one can exit from it.
In view of the recent discussion \cite{eternal} of the exit from
inflation in string theory, and in general on the consistent
quantization of de-Sitter Universes \cite{Witten}, we think that
it is useful to propose new mechanisms 
to exit from inflation. In
our previous work \cite{grace}, we had proposed a mechanism based
on non-critical strings, to exit from inflation. We had considered a
two dimensional model with a tachyon field. The presence of the
tachyon field was crucial for an inflationary phase to go smoothly
to a Robertson-Walker phase. 
Notably, in two-dimensions the 
tachyon field is an excitation of the  `massless' matter multiplet,
and is not associated with an instability of the vacuum.

In the present work we have applied this mechanism to a realistic
four-dimensional model of type $0$ string theory, where 
again the tachyon is stabilized by a RR flux field. 
We modified the $\beta$-functions of the
ten-dimensional theory, assuming that the string theory is
non-critical, which in such theories may be caused by fluctuations of the 
3-brane worlds.
This hypothesis introduces new terms in the
$\beta$-functions of the theory. Physically these terms express
the fact that our non-critical string theory is performing small
oscillations around the conformal point.

The modified $\beta$-functions supplemented with the Curci-Paffuti
equation implied by world-sheet renormalizability, 
have been reduced to four-dimensions. 
Assuming a
homogeneous and spherically symmetric background, we have solved
numerically the resulting equations.
Our solution 
demonstrates  
that the scale factor of the Universe, 
after the initial singularity, enters a
short inflationary phase and then, in a smooth way, goes onto 
a flat
Minkowski spacetime with a linear dilaton for $t \rightarrow \infty$.
The 
fields $\sigma_{1}$ and $\sigma_{2}$ which
parametrise the internal space have an interesting behaviour. The
field $\sigma_{1}$, which sets the scale of the fifth dimension,
during inflation contracts until it reaches a constant value.
After inflation, it maintains this value, until the universe
evolves to an asymptotically flat spacetime
(in the string frame, or 
a linearly expanding, non-accelerating Universe 
in the Einstein frame). The field $\sigma_{2}$
which parametrises the conformally flat five-dimensional space freezes to a
constant value which is much smaller than that of 
the fifth dimension. Thus we see that, in our model,  
a cosmological
evolution may lead to different scales for the extra
dimensions. 

It is important to notice
that the contraction of the extra
dimensions is due to the fact that gravity is very
weak asymptotically in our model. It is also
worthy of stressing
that the possibility of having one large extra
dimension is achieved upon the choice of the RR
flux along that direction. Hence, such a possibility 
arises in our model 
from string non-perturbative effects.
A phenomenologically important feature of the model is that
the vacuum energy, determined by the central-charge deficit,
relaxes to zero asymptotically in a way which is reminiscent
of quintessence models, with the r\^ole of the 
quintessence field played by the dilaton.

The type-$0$ string Universe does not accelerate eternally, 
thus avoiding 
the problem associated with the 
presence of cosmological horizons, 
as far as the definition 
of proper asymptotic states, and hence an S-matrix, 
is concerned. 
A detailed study 
of such issues, as well as of re-heating 
and other 
physical features of inflationary models, 
will be the subject of 
forthcoming work.

We close our discussion 
with an important comment concerning the initial 
singularity of our solution. 
The singularity is a general feature of the equations of
the form (\ref{eqsmotion}). 
Removing the singularity is probably a matter of
a full quantum description of the theory,
which at present is not available. We note 
at this stage, however, the possibility
of deriving smooth cosmological solutions in string theory,
without initial singularities, by including in the action
higher curvature terms (e.g. quadratic 
of Gauss-Bonnet type~\cite{atr}), which are part of the quantum
corrections, in the sense of being generated by 
including string-loop corrections in the effective action.
We plan to return to a systematic study of the effects of
such higher-order 
terms, within our framework, in due course.

\section*{Acknowledgements}

The work of E.P and I.P is partially supported by the NTUA program
Archimedes. G.A.D and B.C.G. would like to acknowledge partial
financial support from the Athens University special account for
research.

\section*{Appendix \\ Dilaton Equations in Liouville Strings} 

In this Appendix we explain in some detail 
why the dilaton field obeys an equation of the form (\ref{neweq}),
in a similar manner with the rest of the background fields/couplings of the 
deformed $\sigma$-model. 
At a $\sigma$-model level, the dilaton field couples
to the world-sheet curvature in the same way as the Liouville
mode, which implies that in general the dilaton 
of the Liouville-dressed theory 
will 
receive contributions from the Liouville mode $\rho$.  
The Liouville mode $\rho$ is viewed as an extra target space dimension,
which has Minkowskian signature in supercritical theories~\cite{aben}.
In our approach, we first treat this extra time dimension as a 
{\it second time}, which, however, is {\it eventually} going to be 
identified
with the existing time-coordinate $t$ of the non-critical string. 

In this sense, the restoration of conformal invariance
implied by the Liouville dressing~\cite{ddk} means that 
in $D+1$ dimensions, with $D$ the original dimensionality 
of the non-critical theory, one would have a conformal 
string. In terms of the ${\tilde \beta}^\Phi$ function 
we would have:
${\tilde \beta}^{\Phi (D+1)} = C^{(D+1)}_{\rm tot} - 10 = 0 $, where 
$C^{(D+1)}_{\rm tot}$ 
is the total 
central charge of the Liouville-dressed theory, and the superscript 
$D+1$ denotes $(D+1)$-dimensional quantities.
To order ${\cal O}(\alpha')$ in a Regge slope $\alpha'$ expansion,
one has for a $d$-dimensional target space:
\be
C^{(d)}_{\rm tot}=d - \frac{3}{2}\alpha'\left[R - 4(\nabla \Phi)^2
+ 4 \nabla^2 \Phi + \frac{1}{4}(\nabla T)^2 + \frac{1}{4}m^2T^2 \right]
\label{cd+1}
\ee 
In general, for non-critical strings
with central-charge deficit $Q^2$, 
one should 
replace $d$ by $3(Q^2 + d^*)$, where $d^*$ is the critical
value of the central charge for the conformal (fixed-point) theory 
($d^*=10$ for superstrings we are dealing with here).  
For convenience we set from now on $\alpha'=2$,
which is the normalization used in (\ref{betafunctions})
and throughout this work.

One now recalls the definition ${\tilde \beta}^{\Phi (D+1)} =
\beta^{\Phi (D+1)} - G^{\mu\nu}\beta^{G(D+1)}_{\mu\nu}$, 
where $\mu,\nu $ span the $D+1$-dimensional spacetime, 
including the Liouville dimension $\rho$.
In Liouville theory, the metric $G_{\rho\rho}=-1$, $G_{\rho M}=0$,  
and hence 
$\beta^{G}_{\rho\rho}=0=\beta^{G}_{\rho M}$, 
which implies that the Liouville 
renormalization does not introduce any extra contributions
to the trace of the graviton $\beta$ function, and therefore to 
the space-time curvature terms in the expression 
for ${\tilde \beta}^\Phi$ (\ref{betafunctions}).
This leaves one with the following condition, expressing
restoration of conformal invariance:
\be
4G^{\rho\rho}\left[(\partial_\rho \Phi)(\partial_\rho \Phi)    
- \partial_{\rho}\partial_{\rho}\Phi 
- \frac{1}{2}G^{MN}(\partial_\rho G_{MN})(\partial_\rho \Phi)  
+ \frac{1}{16}(\partial_\rho T)(\partial_\rho T)\right] + 
{\tilde \beta}^{\Phi (D)} = 0 
\label{dild+1}
\ee
where ${\tilde \beta}^{\Phi (D)}$ 
refers to the
$D$-dimensional parts of the dilaton Weyl anomaly 
coefficient (\ref{betafunctions}), which notably  
include the dilaton field.

As we shall discuss below, the terms 
$\frac{1}{2}G^{MN}\partial_\rho G_{MN}\partial_\rho \Phi $ may be neglected.
There are two ways in which this can be justified. 
First, in our approach~\cite{emn} we treat the Liouville mode $\rho$ as 
a world-sheet renormalization-group scale. Hence terms like 
$\partial_\rho G_{MN}=
\beta^G_{MN}$ and $\partial_\rho \Phi =\beta^\Phi$  
coincide with the world-sheet graviton and dilaton $\beta$-functions. Thus,
such terms are of higher order, as compared with the rest of the terms 
in (\ref{dild+1}),  
if one works in the neighbourhood of a fixed point, which we assume here. 
Alternatively, such quadratic in the $\beta$-functions 
terms can be eliminated by an appropriate 
renormalization-group
scheme choice~\cite{remark}, as we shall discuss later. 

We now assume that the Liouville mode $\rho$ contributions to 
$\Phi (\rho, t, \dots )$ can be represented 
by 
\be
\Phi (\rho, t, \dots ) = -\gamma Q(\rho)\rho +  \varphi (\rho, t, \dots)   
\ee 
where the $\dots$ denote possible dependence on spatial target coordinates, 
$\gamma$ is a numerical constant, to be determined below,  
and  
the term proportional to $Q\rho$ is motivated by the 
Liouville dynamics~\cite{ddk,emn}.
Under this splitting,
and taking into account that ${Q}$ 
does not depend on any other coordinate
except the local renormalization-group scale $\rho$~\cite{emn},   
one may write (\ref{dild+1}) for the supercritical string case 
$G^{\rho\rho}=-1$: 
\be
4{\ddot \varphi} - 4\gamma^2 {Q}^2 + 8\gamma {Q} {\dot \varphi}  
= -{\tilde \beta}^{\Phi} + {\cal O}[(\partial_\rho \Phi)^2,
(\partial_\rho T)^2, {\partial_\rho Q}, (\partial_\rho G)(\partial_\rho \Phi)]    
\label{varp}\ee
where 
${\partial_\rho  Q} \propto (\beta^i)^2$ by virtue of 
the Zamolodchikov $c$-theorem~\cite{zam}, 
extended to the local scale $\rho$~\cite{emn}.

Then, terms of the form 
${\cal O}[(\partial_\rho \varphi)^2,
(\partial_\rho T)^2, {\partial_\rho  Q},
(\partial_\rho G)(\partial_\rho \Phi) ]$ 
can be removed by field redefinitions 
(appropriate renormalization-group scheme choice).
This is possible due 
to the gradient flow property (\ref{flow}) of 
the $\beta$-functions~\cite{remark}, which can be 
shown~\cite{emn} to characterize the Liouville 
problem upon considering~\cite{emn} the Liouville mode as 
a local scale on the world-sheet, appropriate 
for curved-(two-dimensional)-space renormalization~\cite{osborn}.     
Indeed, consider an infinitesimal  
field-redefinition $\delta g^i$ (or 
equivalently $\delta \lambda^i$ for the Liouville-dressed couplings),
where now the set $g^i$ includes the dilaton field. 
This expresses a particular choice of (a class) of renormalization
group schemes. Under this redefinition, the central charge deficit 
$Q^2$, which is assumed to satisfy the gradient flow property 
$\partial_i (Q^2) = {\cal G}_{ij}\beta^j$,  
changes as 
\be
\delta (Q^2) = \delta g^i \partial_i (Q^2) = 
\delta g^i {\cal G}_{ij}\beta^j
\ee
It is then clear that by choosing 
appropriately $\delta g^i \propto \beta^i$,
one may absorb terms of the form 
${\cal O}[(\partial_\rho \varphi)^2,
(\partial_\rho T)^2, {\partial_\rho  Q},
(\partial_\rho G)(\partial_\rho \Phi) ]$   
on the right-hand-side of (\ref{varp})~\cite{remark}. 
Moreover, as already mentioned, 
such terms may also be neglected 
due the fact that, 
in the 
models of `recoil' fluctuations of D3-brane worlds we 
are dealing with here~\cite{kmw,szabo}, 
the induced non-criticality is marginal, 
thereby forcing us to work in the neighbourhood of a 
world-sheet renormalization-group fixed point. 
Hence,
terms quadratic in the $\beta$-functions,
and thus in $\partial_\rho \lambda^i$, are subdominant 
compared to terms linear in such quantities.

With the above in mind, we now observe that,
upon identifying $\rho=2t$, where $t$ is the target time,
fixing $\gamma =1/4 $, and re-expressing the 
terms involving $\varphi$ in terms of $\Phi$,    
we may rewrite (\ref{varp}) in the form (\ref{neweq}),
up to terms (denoted by $\dots$), which may either 
be removed by a renormalization-scheme choice
or are negligible in magnitude for an expansion in the 
neighbourhood of a fixed point we are assuming throughout: 
\be
0 = {\ddot \varphi} + {Q}{\dot \varphi} + 
{\tilde \beta}^\Phi -\frac{1}{4}Q^2 =
{\ddot \Phi} + {Q}{\dot \Phi} + {\tilde \beta}^{\Phi} 
+ \dots = 0 
\label{varpeq}\ee
The overdot denotes differentiation with respect to time $t$. 
Notice that 
the identification of the local world-sheet
renormalization-group scale $\rho$ with the target time
may be understood as a renormalization-group scheme choice~\cite{emn}.

Thus, 
in this way, the dilaton equation (\ref{varpeq}) 
is now expressed as an equation of the form (\ref{neweq}). 
Therefore, by an appropriate scheme choice, one  
was able use the equations (\ref{neweq}) 
for all $\sigma$-model couplings, including the dilaton,
in a way consistent with {\it both} the restoration of 
conformal invariance by the Liouville mode, {\it and} 
the identification of the latter with the target time.

\end{document}